\documentclass{Interspeech}
\usepackage{times}
\usepackage{graphicx}
\usepackage{booktabs}
\usepackage{siunitx}
\usepackage{amsmath}
\usepackage{hyperref}
\usepackage[margin=1in]{geometry}
\usepackage{tabularx,makecell}
\newcommand{\cmark}{\checkmark}
\newcommand{\xmark}{\texttimes}

\interspeechcameraready 

\title{RIR-Mega-Speech: A Reverberant Speech Corpus with Comprehensive Acoustic Metadata and Reproducible Evaluation}

\author[affiliation={1}]{Mandip}{Goswami}

\affiliation{}{Acoustics Researcher}{}
\email{mandipgoswami@gmail.comm}

\usepackage{comment}

\begin{document}

\maketitle

\begin{abstract}
Despite decades of research on reverberant speech, comparing methods remains difficult because most corpora lack per-file acoustic annotations or provide limited documentation for reproduction. We present RIR-Mega-Speech, a corpus of approximately 117.5 hours created by convolving LibriSpeech utterances with roughly 5,000 simulated room impulse responses from the RIR-Mega collection. Every file includes RT60, direct-to-reverberant ratio (DRR), and clarity index ($C_{50}$) computed from the source RIR using clearly defined, reproducible procedures. We also provide scripts to rebuild the dataset and reproduce all evaluation results. 

Using Whisper small on 1,500 paired utterances, we measure 5.20\% WER (95\% CI: 4.69--5.78) on clean speech and 7.70\% (7.04--8.35) on reverberant versions, corresponding to a paired increase of 2.50 percentage points (2.06--2.98). This represents a 48\% relative degradation. WER increases monotonically with RT60 and decreases with DRR, consistent with prior perceptual studies. While the core finding that reverberation harms recognition is well established, we aim to provide the community with a standardized resource where acoustic conditions are transparent and results can be verified independently. The repository includes one-command rebuild instructions for both Windows and Linux environments.
\end{abstract}

\section{Introduction}
Reverberation continues to degrade automatic speech recognition in real-world conditions. The problem is straightforward: late reflections smear temporal structure and reduce the clarity of phonetic cues. Self-supervised models have improved robustness to some channel effects, but rooms with long decay times or weak direct paths still cause errors.

The challenge for research is not just building better models but doing it in a way that others can verify. Many existing reverberant corpora either lack per-file acoustic labels, use proprietary RIRs that cannot be redistributed, or provide limited documentation for reproducing evaluations. This makes it hard to compare results across papers or understand whether improvements generalize.

Our contribution is a corpus with three specific properties. First, every reverberant file has associated RT60, DRR, and $C_{50}$ values computed from the source RIR. Second, we provide complete code to regenerate the audio, compute all metrics, and reproduce the evaluation results in this paper. Third, we report confidence intervals using nonparametric bootstrap at the utterance level and use paired tests where applicable.

We are not claiming algorithmic novelty or surprising acoustic insights. The fact that higher RT60 increases WER is not new. What we offer is a standardized resource where the conditions are documented and the results are checkable. Section 2 reviews related corpora. Section 3 describes the construction process. Sections 4 and 5 present the evaluation setup and results. Section 6 discusses limitations and future directions.

\section{Related Work}

\subsection{Reverberant Speech Corpora}
The REVERB Challenge \cite{kinoshita2016reverb} remains a widely used benchmark. It includes real room recordings and simulated reverberant speech with RT60 ranging from 0.25 to 0.7 seconds. Paired clean and reverberant data are provided, and evaluation scripts are available. However, per-file acoustic metadata is not included in the released annotations, which limits post-hoc analysis by acoustic condition.

The CHiME series \cite{barker2018fifth,watanabe2020chime} has focused on noisy and reverberant multi-channel speech. CHiME-5 and CHiME-6 include real dinner party recordings with substantial reverberation. The acoustic conditions vary naturally but are not systematically annotated with RT60 or DRR per utterance. Simulated training data in some tracks does include controlled reverberation, but the focus is on the challenge task rather than releasing a general-purpose annotated corpus.

AISHELL-4 \cite{fu2021aishell} provides Mandarin meeting speech with reverberation and overlapping speakers. The conference room setting introduces moderate RT60, but again, detailed acoustic parameters are not provided per file. Similar limitations apply to other meeting corpora like AMI \cite{carletta2005ami}.

The Acoustic Echo Cancellation Challenge \cite{sridhar2021icassp} and the Deep Noise Suppression Challenge \cite{reddy2021interspeech} release large sets of RIRs and noise files. These are valuable for data augmentation but are not paired with transcribed speech in a ready-to-use format. Users must perform convolution and manage metadata themselves.

Several VCTK-based reverberant datasets exist \cite{valentini2016investigating,valentini2017noisy}, typically created by convolving VCTK utterances with simulated or measured RIRs. These are often used for speech enhancement evaluation. They provide clean-reverberant pairs but usually lack comprehensive per-file acoustic annotations or systematic coverage analysis.

\subsection{Acoustic Measures in Speech Research}
RT60, introduced by Sabine \cite{sabine1922acoustics}, quantifies how long it takes for sound energy to decay by 60 dB. Schroeder \cite{schroeder1965new} developed the integrated impulse response method, which is now standard for estimating RT60 from measured RIRs. The direct-to-reverberant ratio (DRR) captures the balance between direct sound and reflected energy \cite{zahorik2002direct}. Clarity metrics like $C_{50}$ and $C_{80}$ \cite{reichardt1975definition} are widely used in room acoustics to predict speech intelligibility.

Prior work has established that RT60 above 0.5 seconds typically reduces intelligibility \cite{nabelek1982reverberant}, and that DRR below 0 dB makes speech difficult to understand \cite{bradley2003predictors}. Our evaluation confirms that these trends also appear in ASR error rates with modern neural models.

\subsection{Reproducibility in Speech Research}
Recent discussions at ISCA workshops and in journals like Speech Communication have emphasized the need for reproducible research \cite{mesaros2020sound}. Releasing code is now common, but releasing the exact data and scripts needed to recreate figures and tables is less so. Our approach follows examples from computer vision \cite{russ} and NLP \cite{wang2018glue}, where benchmark datasets come with standardized evaluation protocols and leaderboards.

\section{Dataset Construction}

\subsection{Source Materials}
Clean speech comes from LibriSpeech \cite{panayotov2015librispeech}, specifically the \texttt{dev-clean} and \texttt{test-clean} subsets. We selected roughly 5,200 utterances to provide a range of durations from 1.5 to 36 seconds. This choice prioritizes a well-studied, publicly available corpus with high-quality transcriptions.

The RIR collection comes from RIR-Mega \cite{goswami2025rirmega}, a large-scale corpus of simulated room impulse responses generated using physics-based acoustic simulation methods. RIR-Mega includes responses across diverse room configurations including varied dimensions, absorption coefficients, source-receiver distances, and microphone placements. The simulations cover office spaces, conference rooms, classrooms, auditoriums, and other indoor environments. Our subset contains approximately 5,000 distinct RIRs sampled from the full RIR-Mega collection to provide broad acoustic coverage while maintaining computational feasibility.

We acknowledge that simulated RIRs may not fully capture all complexities of real-world acoustics, such as diffraction around irregular objects, non-uniform surface materials, or time-varying conditions. However, simulation offers several advantages for this work: complete control over acoustic parameters, reproducibility, and the ability to generate large-scale coverage across the RT60-DRR space. The RIR-Mega collection was designed specifically to address limitations in existing RIR datasets by providing systematic coverage of diverse acoustic conditions with verified ground-truth parameters.

We do not claim that this collection is representative of all real-world acoustic environments. Office and conference room geometries are over-represented compared to, say, outdoor or vehicle interiors. The distribution reflects the design of RIR-Mega and our sampling strategy, which prioritized breadth over matching any particular real-world distribution.

\subsection{Convolution and Pairing}
For each clean utterance, we randomly sample up to ten RIRs from the pool and perform time-domain convolution:
\begin{equation}
y[n] = (x * h)[n] = \sum_{k} x[k]\, h[n-k].
\end{equation}
The number of reverberant variants per clean file is not always exactly ten because we exclude RIRs that produce clipping or have problematic metadata. This gives us 53,230 reverberant files totaling 117.5 hours.

Each reverberant file is saved as 16-bit PCM WAV at 16 kHz. We also store a CSV with columns for the clean file ID, the RIR path, and all computed acoustic parameters. This universal metadata file allows filtering and grouping by acoustic condition without loading audio.

\subsection{Acoustic Parameter Computation}
We compute three standard measures for each RIR before convolution.

\textbf{RT60:} We use the Schroeder backward integration method \cite{schroeder1965new}. The energy decay curve is computed from the squared RIR, integrated backward in time, and converted to decibels. We fit a line to the portion of the curve between -5 dB and -35 dB relative to the initial level, then extrapolate to -60 dB. This range avoids early time artifacts and noise floor contamination. The implementation follows ISO 3382-1 \cite{iso3382} recommendations.

\textbf{DRR:} We define a direct-only window of 2.5 ms centered on the first-arrival sample. The direct energy is the sum of squared samples within this window. Reverberant energy is the sum of squared samples outside the window. DRR in decibels is:
\begin{equation}
\mathrm{DRR} = 10 \log_{10} \frac{\sum_{t \in \text{direct}} h^2[t]}{\sum_{t \notin \text{direct}} h^2[t]}.
\end{equation}
This definition differs from some prior work that includes early reflections (up to 50 ms) in the direct component. We chose a narrow window to isolate the true direct path. This may not align with perceptual definitions where early reflections contribute to clarity, and it can produce very low DRR values when the direct peak is weak. We acknowledge this as a limitation and plan to add alternative DRR definitions in future releases.

\textbf{$C_{50}$:} The clarity index at 50 ms is:
\begin{equation}
C_{50} = 10 \log_{10} \frac{\sum_{t=0}^{50\,\text{ms}} h^2[t]}{\sum_{t>50\,\text{ms}} h^2[t]}.
\end{equation}
This is a standard metric in room acoustics and is often used to predict speech intelligibility \cite{bradley2003predictors}.

We also compute a loudness proxy (RMS in dB) and duration in seconds for each reverberant file. These are used in normalization experiments.

\subsection{Dataset Splits and Limitations}

We define train, development, and test splits stratified by speaker to prevent speaker overlap across partitions. The universal metadata CSV includes a split column with values train, dev, or test. Table 1 shows the distribution: 43,660 files (82.0\%) for training, 4,620 files (8.7\%) for
development, and 4,950 files (9.3\%) for testing. Separate CSV files (train.csv, dev.csv, test.csv) are also provided in the metadata directory for convenience. The RIR selection process is not stratified by acoustic parameters. We sample uniformly from the available pool, which means coverage is uneven. Figure 3 shows that some RT60-DRR regions have many samples while others have few. A stratified sampling approach would improve balance but would require a larger RIR pool or targeted simulation. 

The clean speech is limited to LibriSpeech, which means read English from audiobooks. Spontaneous speech, non-native accents, and other languages are not covered. We chose LibriSpeech for reproducibility and because transcriptions are reliable, but this limits ecological validity.

\section{Corpus Statistics and Coverage}

Table~\ref{tab:split_summary} summarizes the metadata for all reverberant files. The mean RT60 is 0.44 seconds with a range from 0.09 to 1.51 seconds. Mean DRR is 3.32 dB but the range is large due to our narrow direct-window definition. Some RIRs have very weak direct peaks, which produces negative DRR values below -100 dB. These files are outliers but are included for completeness.

The corpus is divided into train, development, and test splits with 43,660, 4,620, and 4,950 files respectively. Splits are stratified by speaker following LibriSpeech conventions: all utterances from a given speaker appear in only one split. This ensures that speaker-specific characteristics do not leak between training and evaluation. The acoustic parameter distributions (RT60, DRR, C50) are similar across splits, though we did not explicitly balance by acoustic condition.
\begin{table}[t]
\centering
\caption{Corpus statistics. Splits: 82\% train, 8.7\% dev, 9.3\% test.}
\label{tab:split_summary}
\small
\begin{tabular}{lccccc}
\toprule
Metric & Mean & Median & Std & Min & Max \\ 
\midrule
Duration (s) & 7.96 & 6.52 & 4.94 & 1.52 & 36.07 \\
RT60 (s) & 0.44 & 0.36 & 0.25 & 0.09 & 1.51 \\
DRR (dB) & 3.32 & 6.58 & 22.11 & -141.96 & 30.77 \\
\bottomrule
\end{tabular}
\\[0.5em]
\small Total hours: 117.68 \quad Total files: 53,230
\end{table}

\begin{figure}[t]
  \centering
  \fbox{\includegraphics[width=0.9\linewidth]{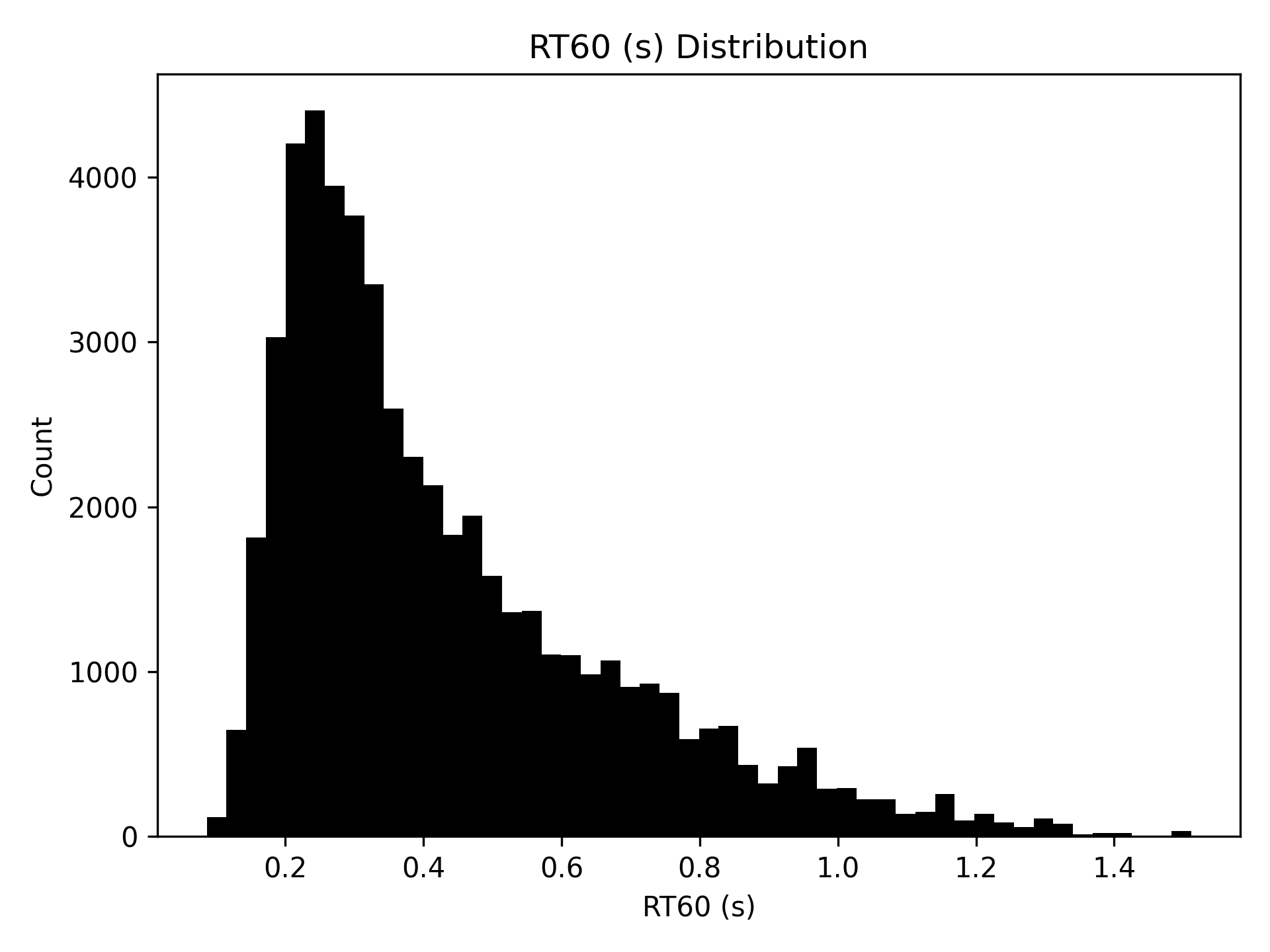}}
  \caption{RT60 distribution across all reverberant files. Most files fall between 0.2 and 0.8 seconds.}
  \label{fig:hist_rt60}
\end{figure}

Figure~\ref{fig:hist_rt60} shows the RT60 distribution. There is a mode around 0.3 to 0.4 seconds, which corresponds to typical office and classroom sizes. The tail extends to 1.5 seconds, which covers larger halls but is not representative of, say, cathedrals or concert halls.

\begin{figure}[t]
  \centering
  \fbox{\includegraphics[width=0.9\linewidth]{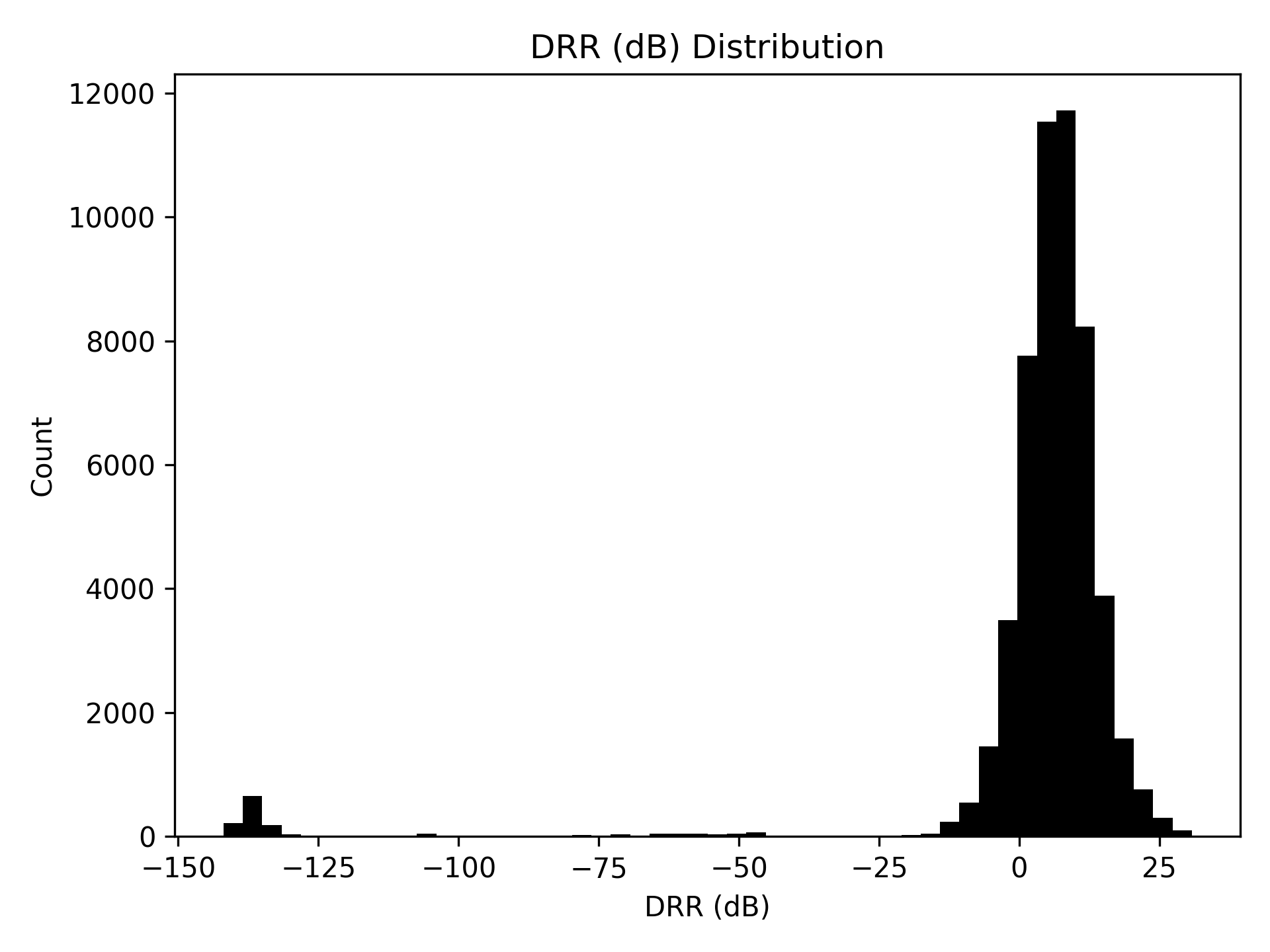}}
  \caption{DRR distribution using a 2.5 ms direct-only window. The long tail toward negative values reflects weak direct arrivals in some simulated RIRs.}
  \label{fig:hist_drr}
\end{figure}

Figure~\ref{fig:hist_drr} shows the DRR distribution. Most files have DRR between -5 and +15 dB. The long tail toward very negative values occurs when the direct peak is weak relative to early reflections, which can happen in simulation for certain source-receiver geometries. We include these files because they represent challenging acoustic conditions, though users should be aware that such extreme values may be less common in typical real rooms.

\begin{figure}[t]
  \centering
  \fbox{\includegraphics[width=\linewidth]{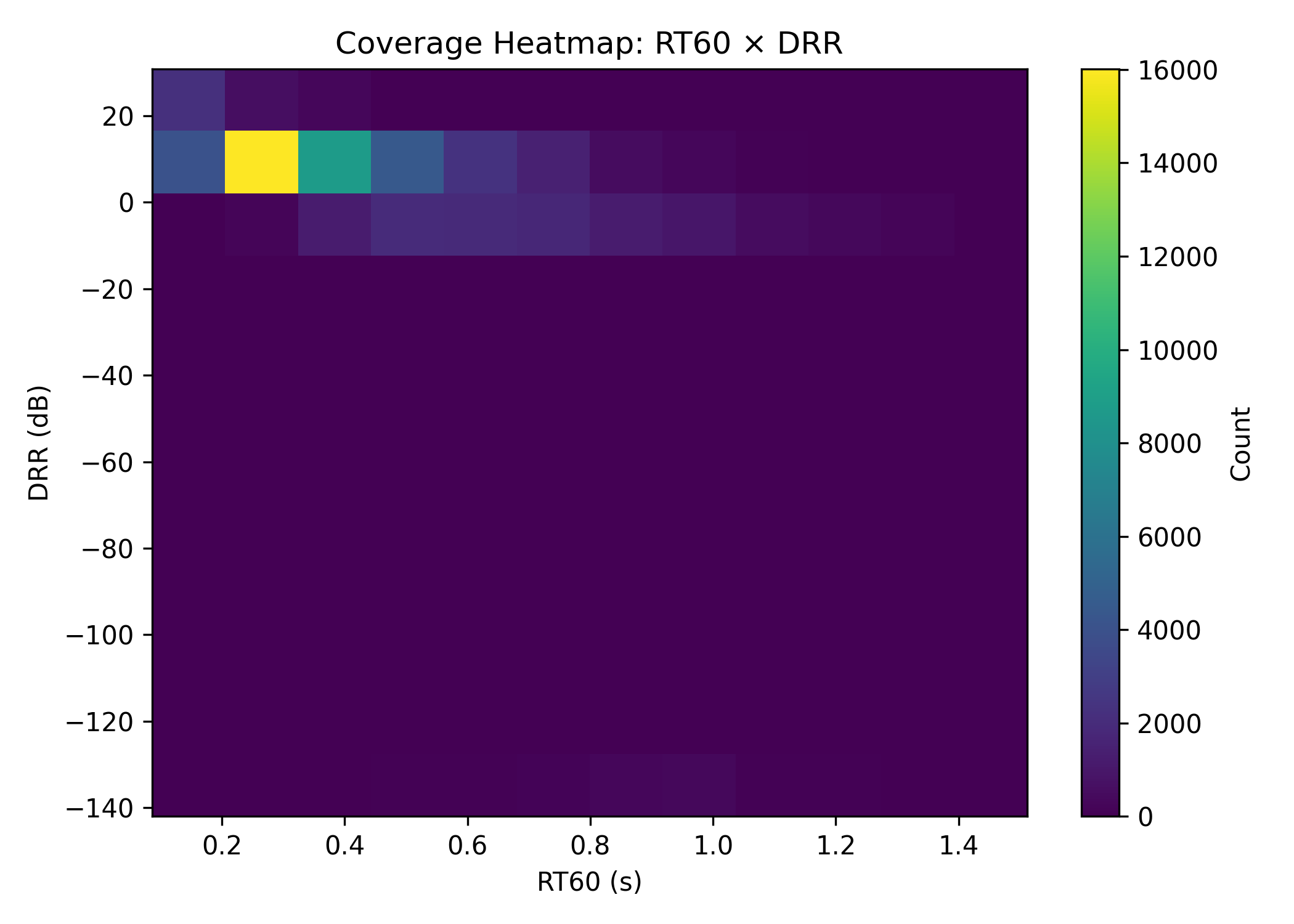}}
  \caption{Coverage heatmap (RT60 vs DRR). Darker cells indicate more files. Coverage is uneven due to non-stratified sampling.}
  \label{fig:coverage}
\end{figure}

Figure~\ref{fig:coverage} is a heatmap of file counts in the RT60-DRR plane. Coverage is densest in the 0.2 to 0.6 second RT60 range with DRR between 0 and 10 dB. Cells with high RT60 and very low DRR have few samples. This reflects the composition of the RIR pool and the fact that extreme combinations are rare in typical rooms.

\begin{figure}[t]
  \centering
  \fbox{\includegraphics[width=\linewidth]{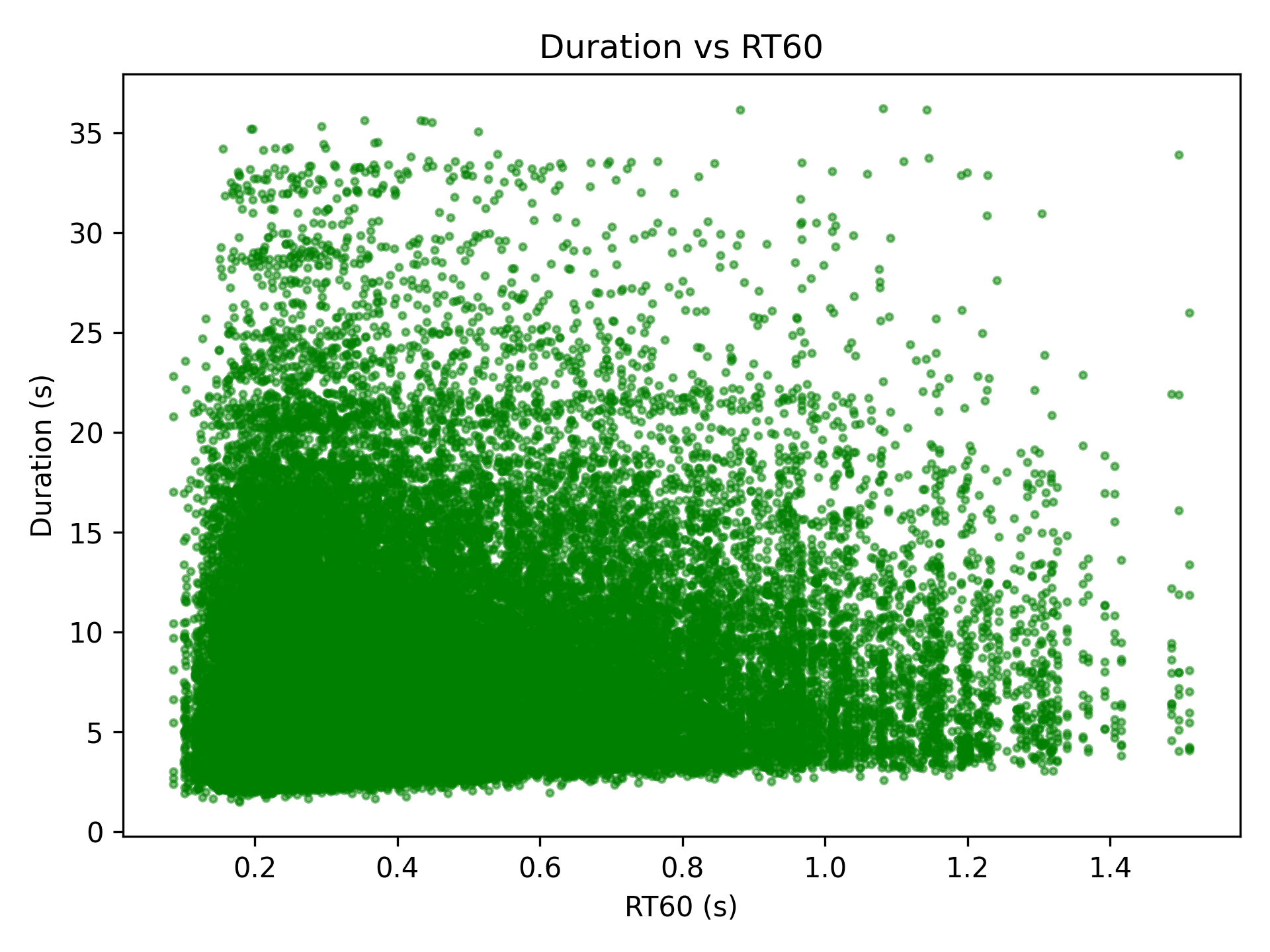}}
  \caption{Duration vs RT60. Long and short utterances are distributed across RT60 bins, reducing confounding in WER analysis.}
  \label{fig:dur_vs_rt60}
\end{figure}

Figure~\ref{fig:dur_vs_rt60} plots duration against RT60 for all files. There is no strong correlation, which means long utterances are not systematically assigned to high or low RT60 conditions. This reduces the risk that duration confounds acoustic trends in WER analysis.

\begin{figure}[t]
  \centering
  \fbox{\includegraphics[width=\linewidth]{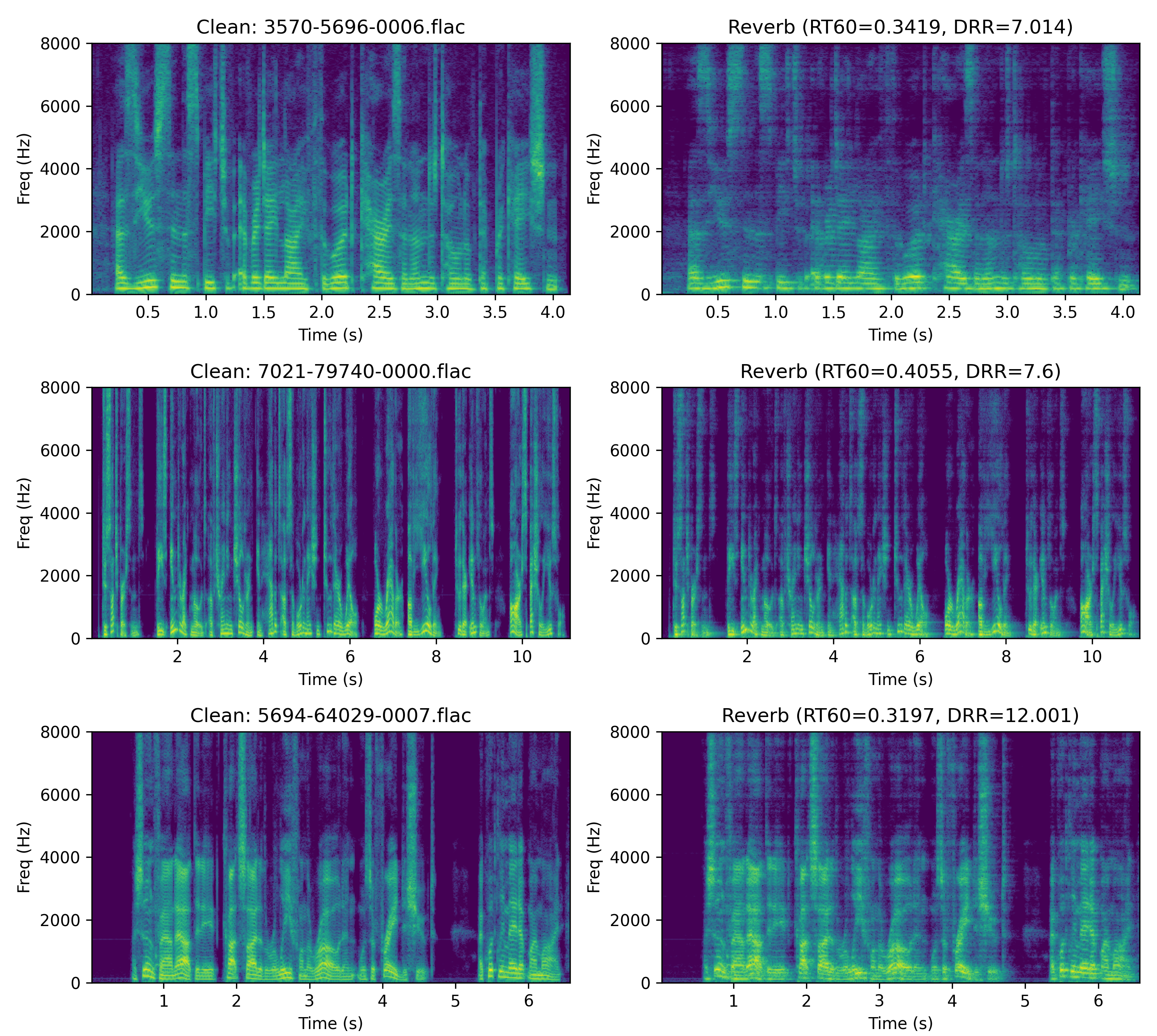}}
  \caption{Spectrogram comparison for three utterances. Left column: clean. Right column: reverberant. Smearing of formant structure is visible.}
  \label{fig:spec_grid}
\end{figure}

Figure~\ref{fig:spec_grid} shows spectrograms for three utterances in clean and reverberant conditions. Energy smearing and the buildup of late reflections are visible. This is included as a qualitative check that the convolution process produces expected results.

\section{Evaluation Setup}

\subsection{Model Selection}
We evaluate using Whisper small \cite{radford2022whisper}. Whisper was trained on a large multilingual dataset and has been shown to be relatively robust to domain shift, which makes it a reasonable starting point. We do not claim this is the best available model, but it is widely used and publicly available.

For Whisper, we use the default beam search with beam size 5. We do not tune any hyperparameters. Decoding uses the English transcription task. We apply the same normalization to reference and hypothesis text (lowercase, remove punctuation) before computing WER.

\subsection{Evaluation Sets}

We construct a paired clean-reverberant set by selecting 1,500 utterances from the test split. For each utterance ID, we decode both the clean file and one randomly chosen reverberant variant. This allows paired statistical tests, which remove between-utterance variance and increase power. All evaluation results in this paper use only test split data to avoid any overlap with potential training scenarios. 
For ablation experiments (loudness normalization and additive noise), we use a smaller set of 500 utterances from the test split due to computational cost. These are a subset of the 1,500 and are chosen to have a range of RT60 and DRR values. 
The train and dev splits are provided for researchers who wish to train dereverberation or robust ASR models. We do not provide baseline results on the dev set in this paper, as our focus is on corpus characterization rather than model development. 
\subsection{Statistical Methods}
For all WER estimates, we report nonparametric bootstrap 95\% confidence intervals. We resample utterances with replacement $B=2000$ times, compute WER on each resample, and take the 2.5th and 97.5th percentiles of the bootstrap distribution. This does not assume normality and is appropriate for the skewed distribution of per-utterance WER.

For the clean versus reverberant comparison, we compute the paired difference in WER for each utterance, then bootstrap the mean difference. This paired test is more powerful than comparing independent samples because it controls for utterance difficulty.

For trends versus acoustic parameters, we bin RT60 and DRR using fixed edges: RT60 bins at [0.2, 0.4, 0.6, 0.8, 1.0, 1.2] seconds; DRR bins at [-10, -5, 0, 5, 10, 15] dB. Within each bin we compute WER and its bootstrap CI. We do not adjust for multiple comparisons because we are not testing hypotheses but rather describing trends.

\section{Results}

\subsection{Clean vs Reverberant: Paired Baseline}
Table~\ref{tab:baselines} shows the main result. On the 1,500 paired utterances, Whisper small achieves 5.20\% WER (95\% CI: 4.69 to 5.78) on clean speech and 7.70\% WER (7.04 to 8.35) on reverberant versions. The paired difference is 2.50 percentage points (2.06 to 2.98), which is a 48\% relative increase. All utterances contribute to the mean, so this includes both easy and hard cases.

\begin{table}[t]
\centering
\caption{Whisper small WER on paired clean and reverberant utterances (N=1,500). Bootstrap 95\% CIs in parentheses.}
\label{tab:baselines}
\small
\begin{tabular}{lc}
\toprule
Condition & WER (\%) \\ 
\midrule
Clean & 5.20 (4.69--5.78) \\
Reverberant & 7.70 (7.04--8.35) \\
\midrule
Paired $\Delta$WER & +2.50 (+2.06--+2.98) \\
Relative increase & +48.2\% \\
\bottomrule
\end{tabular}
\end{table}

\begin{figure}[t]
  \centering
  \fbox{\includegraphics[width=\linewidth]{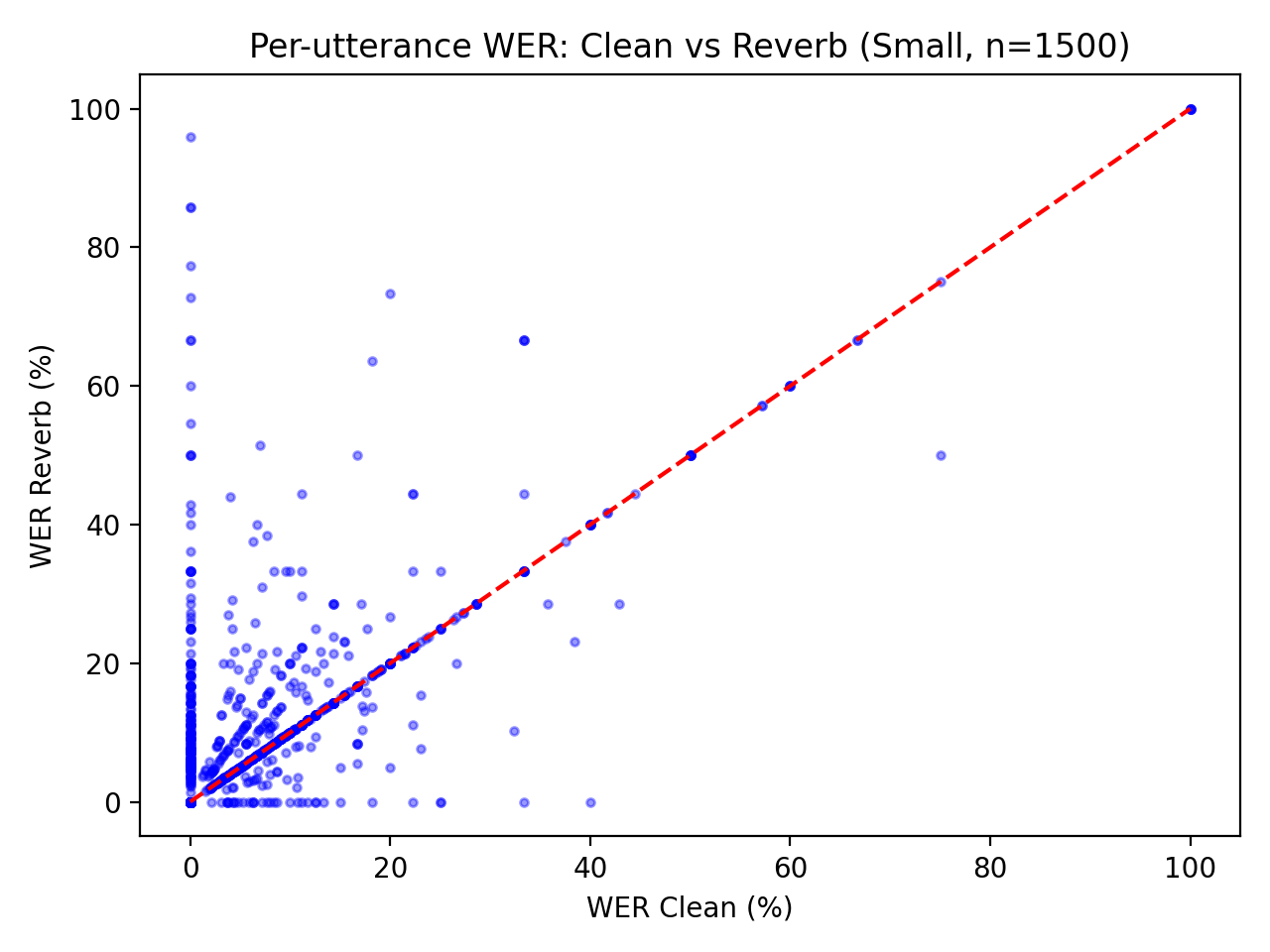}}
  \caption{Per-utterance WER: clean vs reverberant (N=1,500). Most points above the diagonal indicate reverberation increases errors.}
  \label{fig:paired_scatter}
\end{figure}

Figure~\ref{fig:paired_scatter} plots per-utterance WER for clean versus reverberant. Most points lie above the diagonal, meaning reverberation increases errors. A few points lie below the diagonal, which could be noise or cases where reverberation happens to mask artifacts in the clean recording. The scatter is fairly wide, which reflects the range of acoustic conditions and utterance difficulties.

\subsection{Trends with RT60 and DRR}
\begin{figure}[t]
  \centering
  \fbox{\includegraphics[width=\linewidth]{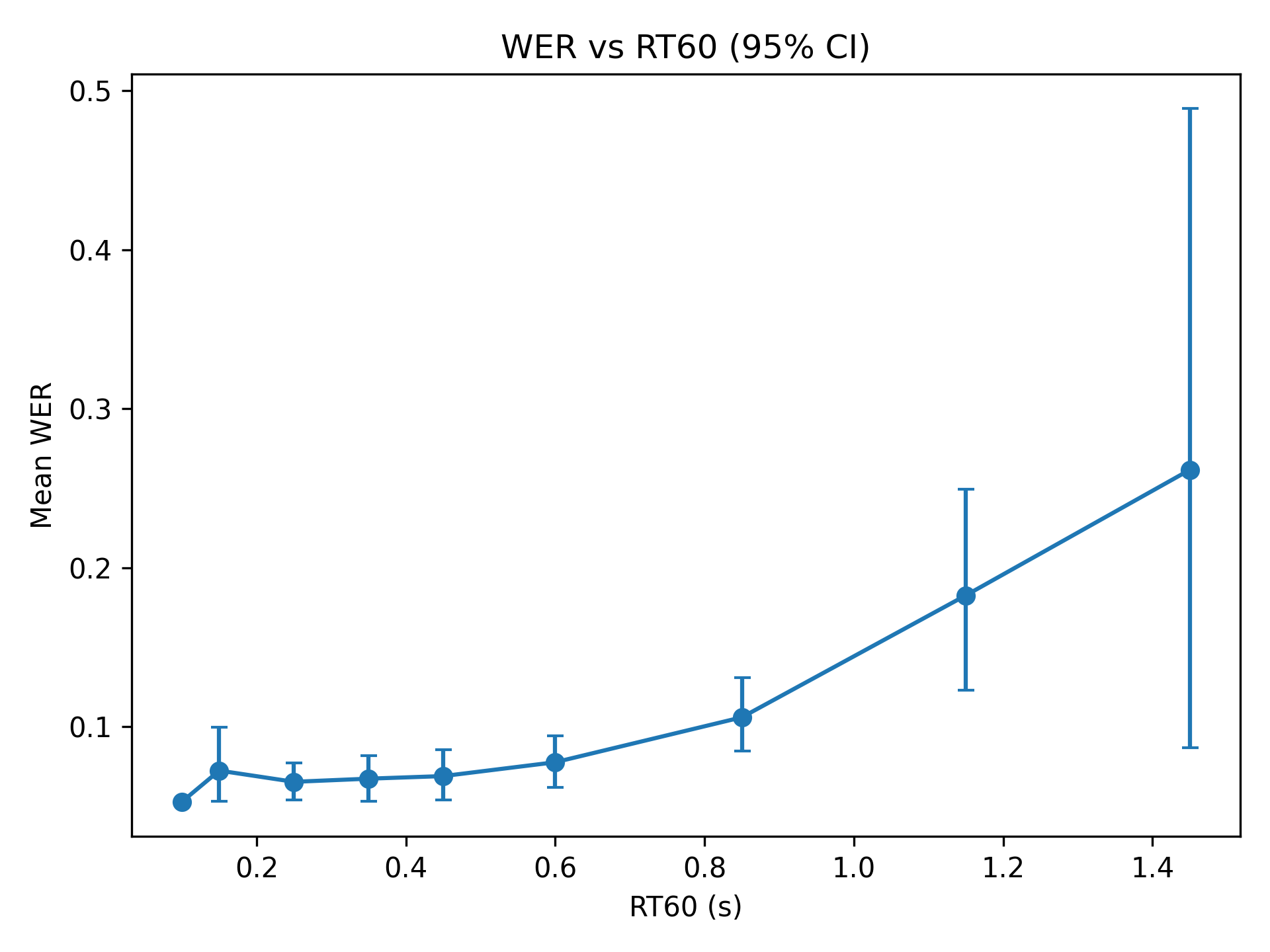}}
  \caption{WER vs RT60 with bootstrap 95\% CIs. WER increases with RT60 as expected. Error bars show confidence intervals.}
  \label{fig:wer_rt60}
\end{figure}

Figure~\ref{fig:wer_rt60} shows WER versus RT60 in fixed bins. WER increases from about 6\% at RT60 = 0.2--0.4 seconds to about 10\% at RT60 = 1.0--1.2 seconds. The trend is monotonic and the confidence intervals do not overlap for extreme bins, which suggests the effect is real. This is consistent with prior studies on intelligibility \cite{nabelek1982reverberant}.

\begin{figure}[t]
  \centering
  \fbox{\includegraphics[width=\linewidth]{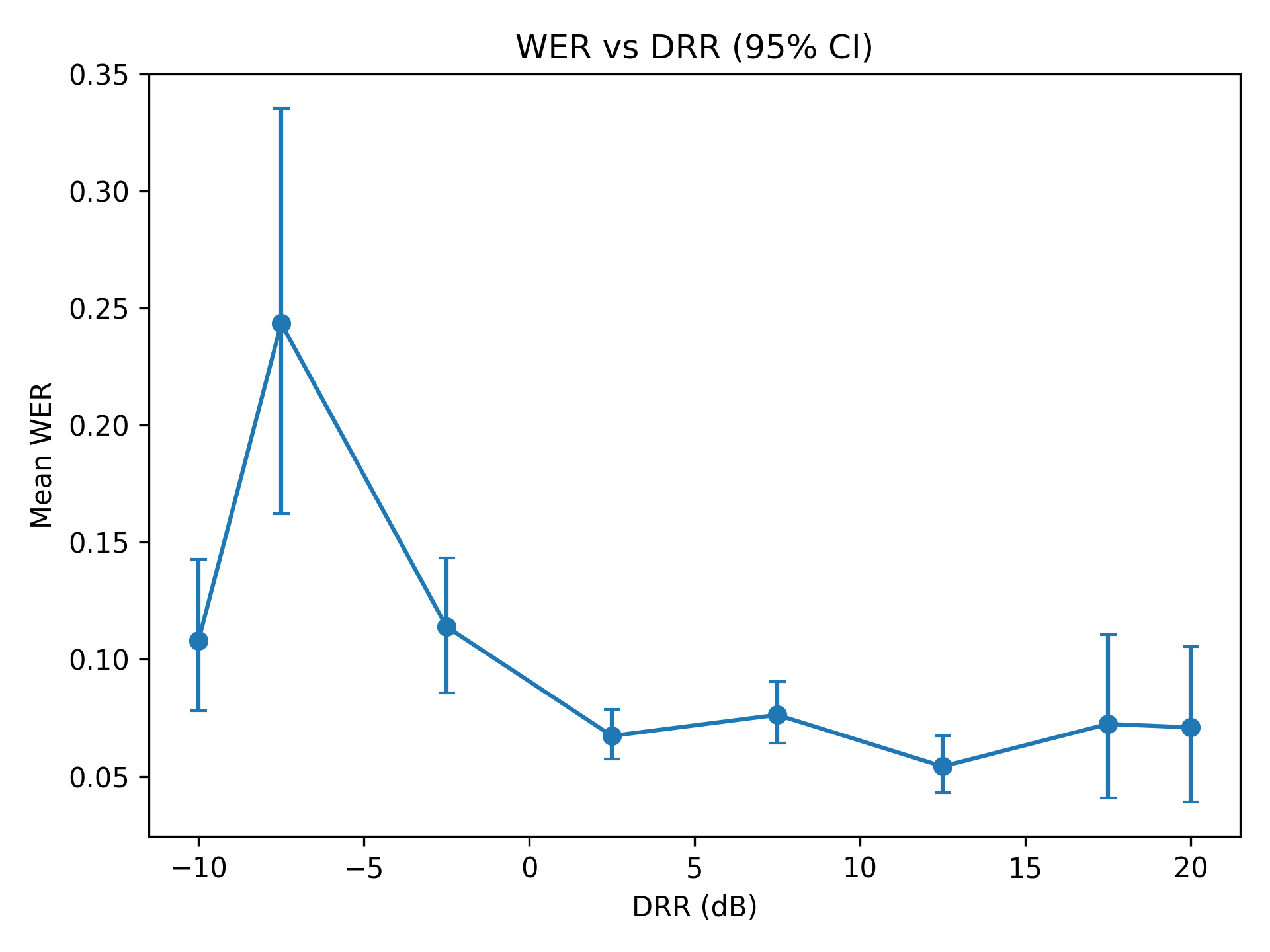}}
  \caption{WER vs DRR with 95\% CIs. Higher DRR (stronger direct path) reduces errors.}
  \label{fig:wer_drr}
\end{figure}

Figure~\ref{fig:wer_drr} shows WER versus DRR. WER decreases as DRR increases, which means better direct-to-reverberant ratio improves recognition. The effect is stronger at low DRR (below 0 dB) where direct energy is weak. At high DRR (above 10 dB), WER plateaus near the clean-speech level.

\begin{figure}[t]
  \centering
  \fbox{\includegraphics[width=\linewidth]{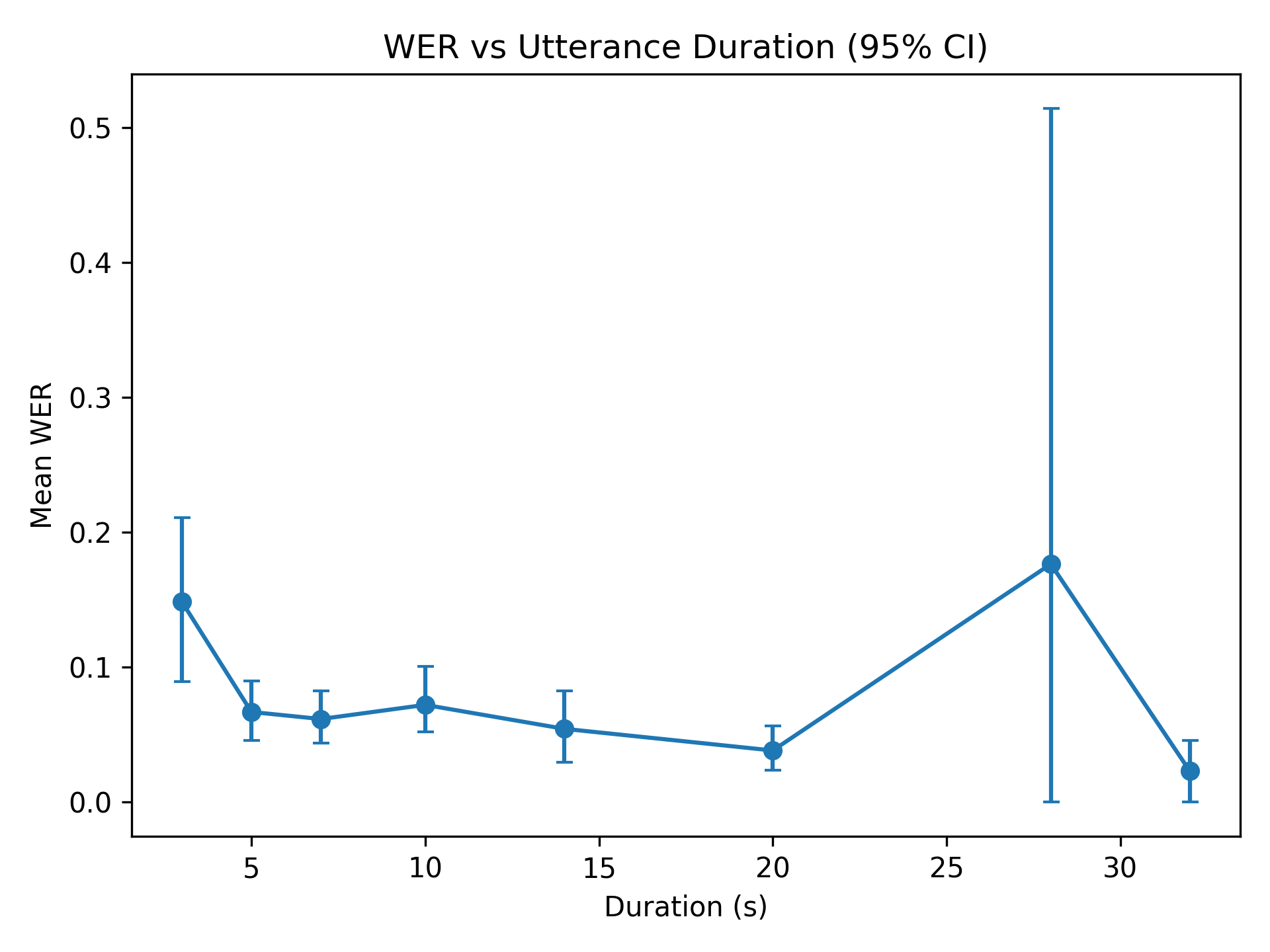}}
  \caption{WER vs duration with 95\% CIs. Duration has a weak effect after controlling for acoustics.}
  \label{fig:wer_duration}
\end{figure}

Figure~\ref{fig:wer_duration} shows WER versus duration. There is a weak upward trend, but the effect is smaller than for RT60 or DRR. This suggests that once acoustic conditions are accounted for, longer utterances are not inherently harder. This is expected because Whisper processes audio in chunks and does not have a strong length penalty.

\begin{figure}[t]
  \centering
  \fbox{\includegraphics[width=\linewidth]{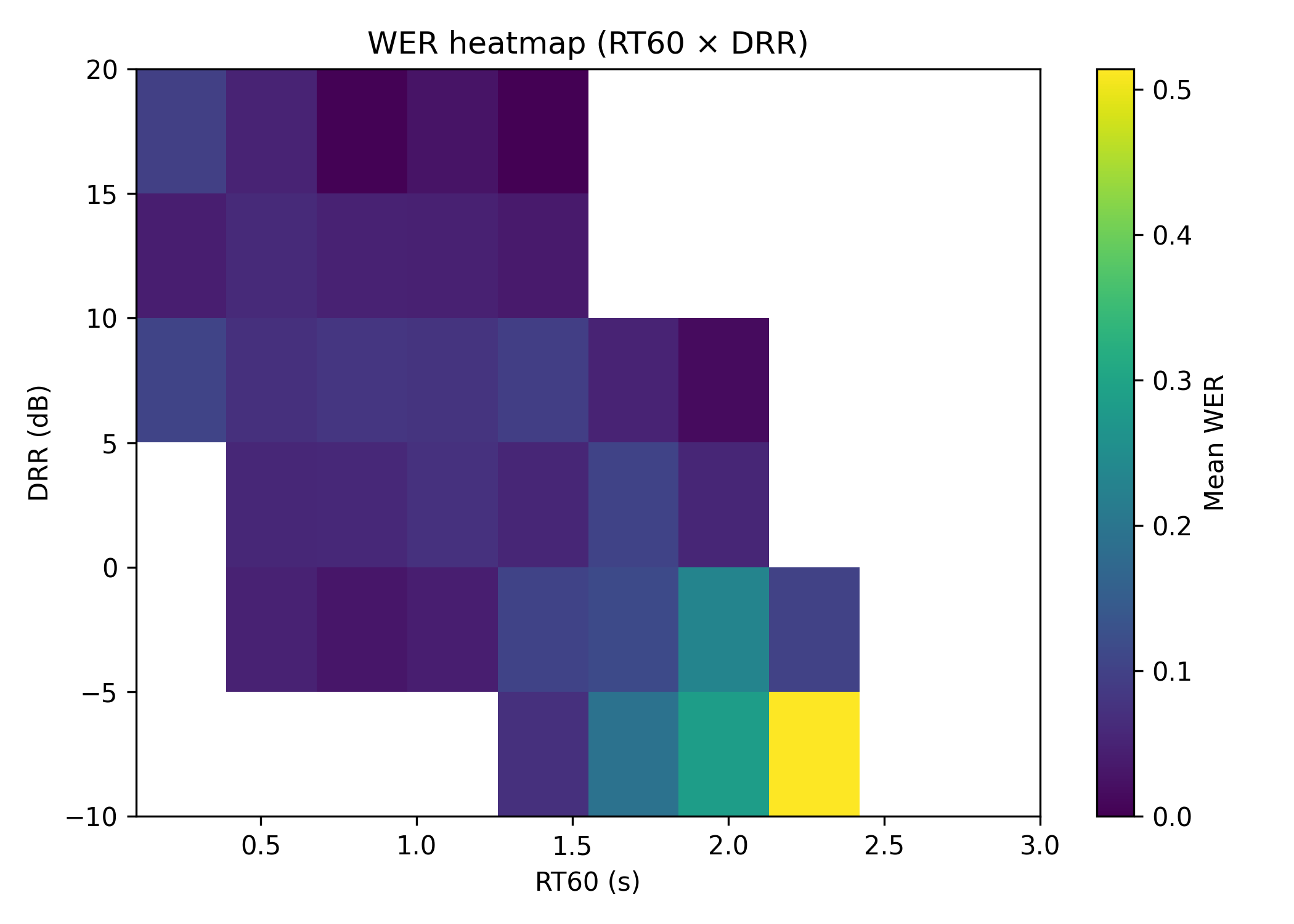}}
  \caption{WER heatmap (RT60 vs DRR). Darker regions indicate higher WER. The worst conditions are high RT60 with low DRR.}
  \label{fig:wer_heatmap}
\end{figure}

Figure~\ref{fig:wer_heatmap} is a heatmap of WER in the RT60-DRR plane. The highest errors occur in the bottom-right region (high RT60, low DRR). The lowest errors are in the top-left (low RT60, high DRR). This visualization confirms that the two acoustic dimensions interact: a room with both long decay and weak direct path is particularly challenging.

\subsection{Ablations: Loudness Normalization and Additive Noise}
On a 500-utterance subset, we test two modifications. First, we normalize using RMS-based loudness (target -20 dB RMS). This gives 8.00\% WER (95\% CI: 6.57 to 9.58), which is close to the baseline reverberant result of 7.70\%. The confidence intervals overlap, so we cannot conclude that normalization changes difficulty. This suggests that loudness variation is not a major factor in this corpus.

Second, we add white noise at a signal-to-noise ratio between 10 and 15 dB (chosen uniformly at random for each file) after loudness normalization. This gives 30.95\% WER (95\% CI: 27.22 to 34.87). The large increase is expected: at these SNRs, additive noise dominates the error budget. This shows that the model is sensitive to noise but does not tell us much about reverberation specifically. We include it as a sanity check.

\subsection{Error Analysis}

We rank all reverberant utterances by per-utterance WER and inspect the top 25 hardest cases. These files have WER above 50\% and are concentrated in conditions with RT60
above 0.8 seconds and DRR below -5 dB. We manually lis- tened to eight of these high-error utterances and compared the Whisper output to the reference transcription. In these
examples, errors included substitutions of phonetically similar consonants (e.g., "sit" recognized as "zit", "bat" as "bad") and deletion of unstressed function words ("the",
"a", "to"). In two cases the reference transcription itself contained disfluencies or partial words that are difficult to recognize even in clean audio, which contributes to inflated WER scores.

We save the complete ranked list to hardest\_utterances\_top25.csv so others can perform their own listening tests. A systematic phonetic analysis would require forced alignment at the phoneme level, which we have not completed. Such an analysis could quantify which phoneme classes (fricatives, stops, vowels) are most affected by specific acoustic conditions and is planned as future work.

\section{Discussion and Limitations}

\subsection{What This Corpus Provides}
RIR-Mega-Speech is intended as a standardized evaluation resource with transparent acoustic metadata. It is not a new benchmark in the sense of introducing a novel task or metric. The value is in the combination of scale, per-file annotations, and reproducibility. Researchers who want to test dereverberation methods or robust ASR models can use this corpus to report results that others can verify.

The one-command rebuild process (using PowerShell on Windows or bash on Linux) is designed to lower the barrier to reproduction. All scripts that generate audio, compute metrics, and produce figures are included. This does not guarantee that results will be identical on different hardware or software versions, but it makes the process explicit.

\subsection{What This Corpus Does Not Provide}

The most significant limitation is that all RIRs are simulated using physics-based methods rather than measured from real rooms. While this provides reproducibility and systematic parameter control, simulated RIRs may not capture all complexities of real rooms such as non-uniform surface scattering, furniture effects, or HVAC noise. Researchers requiring validation on real-world recordings should supplement evaluation with corpora like REVERB or CHiME that include measured room responses.
The acoustic coverage is uneven. Some RT60-DRR regions have hundreds of files while others have fewer than ten. The DRR definition we use (2.5 ms direct-only window) is non-standard. Most room acoustics work includes early reflections (up to 50 ms) in the direct component because they contribute to intelligibility. Our narrow window was chosen to isolate the true direct path, but it produces very low DRR values in some cases and may not align with perceptual relevance. We plan to add alternative DRR definitions in future versions.

\subsection{Comparison with Existing Corpora}
Table~\ref{tab:related_features} compares RIR-Mega-Speech with other reverberant corpora on several practical dimensions. The main difference is that we provide per-file acoustic annotations in a single CSV and include one-command rebuild scripts. REVERB and CHiME provide paired clean-reverberant data but do not ship per-file RT60/DRR annotations. DNS Challenge provides RIRs but not convolved speech. VCTK-based sets vary widely depending on the release.

We are not claiming that RIR-Mega-Speech is better than these corpora in all respects. REVERB and CHiME have real recordings with natural room characteristics that our simulated RIRs may not fully capture. The advantage of our approach is systematic acoustic coverage, documented ground-truth parameters, and reproducibility. For validation of dereverberation or robust ASR methods, we recommend evaluation on both simulated corpora like ours and real-world recordings like REVERB to assess generalization.

\begin{table}[t]
\centering
\caption{Comparison of corpora on practical features.}
\label{tab:related_features}
\scriptsize
\setlength{\tabcolsep}{2.5pt}
\begin{tabular}{lcccc}
\toprule
Corpus & \makecell{RT60/\\DRR} & \makecell{Paired} & \makecell{Rebuild\\script} & \makecell{Boot-\\strap} \\
\midrule
REVERB & \xmark & \cmark & varies & \xmark \\
CHiME-5/6 & \xmark & \cmark & varies & \xmark \\
DNS RIRs & varies & \xmark & varies & \xmark \\
AISHELL & \xmark & \cmark & \xmark & \xmark \\
VCTK-rev & varies & \cmark & varies & \xmark \\
\midrule
\textbf{Ours} & \textbf{\cmark} & \textbf{\cmark} & \textbf{\cmark} & \textbf{\cmark} \\
\bottomrule
\end{tabular}
\end{table}

\subsection{Future Directions}
We plan several extensions. First, expand the RIR pool to cover more extreme conditions (RT60 above 2 seconds, outdoor environments, moving sources). Second, add alternative DRR definitions and perceptually motivated metrics like speech transmission index (STI). Third, evaluate more models including dereverberation methods and recent self-supervised architectures. Fourth, release a smaller "RIR-Mega-Lite" subset (10 to 20 hours) for fast prototyping. 
We also hope to add support for languages other than English by convolving other corpora (e.g., Common Voice, AISHELL) with the same RIR pool. This would allow cross-lingual robustness studies. 
\section{Reproducibility Details}
Table~\ref{tab:repro_card} lists the software and hardware environment used to generate the results in this paper. We developed the scripts on Windows 10 with PowerShell 7 but also tested on Ubuntu 20.04 with bash. The main dependencies are Python 3.10, PyTorch, torchaudio, transformers, and librosa. GPU is used for Whisper inference but not required for RIR metric computation or convolution.

\begin{table}[t]
\centering
\caption{Reproducibility environment details.}
\label{tab:repro_card}
\scriptsize
\setlength{\tabcolsep}{3pt}
\begin{tabular}{ll}
\toprule
\textbf{Item} & \textbf{Detail} \\
\midrule
OS & Win 10/11 or Ubuntu 20.04 \\
Shell & PowerShell 7 / bash \\
Python & 3.10 \\
Libraries & torch, torchaudio, \\
 & transformers, librosa \\
GPU & NVIDIA 24 GB (optional) \\
RAM & 64 GB recommended \\
Storage & SSD, 200 GB free \\
Seeds & 42 (sampling); 0 (boot) \\
Bootstrap & $B=2000$ utterances \\
RT60 bins & [0.2, 0.4, 0.6, 0.8, \\
 & 1.0, 1.2] seconds \\
DRR bins & [-10, -5, 0, 5, 10, 15] dB \\
Eval N & 1,500 (main); 500 (abl) \\
Build time & 2--3 h (16 cores) \\
Eval time & 1--2 h (1 GPU) \\
Disk & $\sim$117.5 h audio \\
Command & \texttt{make\_all.ps1/.sh} \\
\bottomrule
\end{tabular}
\end{table}

Runtime to build the corpus (convolve 53,230 pairs) is about 2 to 3 hours on a 16-core CPU. This is I/O bound due to reading and writing WAV files. Evaluation with Whisper small on 1,500 files takes about 1 to 2 hours on a single GPU with 24 GB VRAM. Bootstrap resampling for confidence intervals adds negligible time (under 5 minutes) because it operates on cached WER values.

The repository is organized as follows: \texttt{data/clean\_speech} contains LibriSpeech inputs, \texttt{data/rirmega\_rirs} contains RIRs, \texttt{outputs/rirmega\_speech\_out} contains generated audio and the universal metadata CSV, \texttt{outputs/figures} contains all plots, and \texttt{scripts} contains build and evaluation utilities. Running \texttt{make\_all.ps1} on Windows or \texttt{make\_all.sh} on Linux executes the full pipeline.

\section{Planned Releases}
We plan to release the following artifacts with separate DOIs to support different use cases:

\begin{enumerate}
\item \textbf{RIR-Mega-Speech:} Full corpus with universal metadata CSV.
\item \textbf{RIR-Mega-Lite:} 10--20 hour subset for fast benchmarking.
\item \textbf{RIR-Mega-Eval:} Evaluation scripts and baseline result CSVs.
\item \textbf{RIR-Mega-Metrics:} Standalone library for computing RT60, DRR, $C_{50}$.
\end{enumerate}

Licenses will follow the source materials: LibriSpeech is CC BY 4.0, and simulated RIRs from RIR-Mega are released under MIT (see \cite{goswami2025rirmega} for details). Derived audio (convolved files) can be redistributed under the same terms as the clean source.

\section{Conclusion}
We have presented RIR-Mega-Speech, a corpus of reverberant speech with comprehensive acoustic annotations and reproducible evaluation scripts. The corpus uses simulated RIRs from the RIR-Mega collection \cite{goswami2025rirmega}, which provides systematic acoustic coverage with verified ground-truth parameters. The main contribution is not algorithmic novelty but a standardized resource where acoustic conditions are documented and results can be independently verified. Whisper small shows a 48\% relative increase in WER under reverberation, and error rates increase monotonically with RT60 and decrease with DRR. These trends are expected but are now quantified with confidence intervals on a corpus that others can rebuild.

The corpus includes train, development, and test splits stratified by speaker, though acoustic coverage remains uneven and model comparisons are limited to Whisper. We are releasing it with the expectation that future versions will expand acoustic coverage and baseline diversity. We hope it serves as a useful baseline for research on robust ASR and dereverberation. This release documents version v1.0 of the corpus; future releases may revise acoustic metric definitions and expand coverage while preserving backward compatibility.

\section{Acknowledgements}
We thank colleagues who reviewed the scripts and suggested diagnostic plots including the duration-RT60 scatter and side-by-side spectrograms.

\end{document}